\documentstyle[12pt,epsfig]{article}
\begin{document}
\noindent

\begin{center}
{\Large {\bf Attractor Solutions in Interacting Dark Energy Models}}\\
\vspace{2cm}
 ${\bf Yousef~Bisabr}$\footnote{e-mail:~y-bisabr@sru.ac.ir}\\
\vspace{.5cm} {\small{Department of Physics, Shahid Rajaee Teacher
Training University,
Lavizan, Tehran 16788, Iran}}\\
\end{center}
\vspace{1cm}
\begin{abstract}
We investigate a cosmological model in which dark energy, represented by a quintessential scalar
field, is coupled to a dark-matter perfect fluid in the spatially flat Friedmann-Robertson-Walker Universe. This allows an energy exchange in the dark sector which could happen both at early times before recombination era or at late times. We use the coupling function $Q=\gamma\rho_{dm}\dot{\varphi}$ which is induced by conformal transforming scalar-tensor and $f(R)$ gravity theories to Einstein frame. It is argued that there is a connection between this coupling function and $Q\propto \rho_{dm}H$. A dynamical analysis is used to show that there are early- and late-time attracting solutions for which the system evolves for a wide range of initial conditions. These attractors generalize the scaling solutions which have been already found in the non-interacting case.

\end{abstract}

\section{Introduction}
There are a large number of evidences which confirm that $\Lambda$CDM is a successful model in describing cosmological observations. The model allows us to describe the expansion history of the Universe from the beginning until now. One important quantity in this description is $H_0$, the present value of the Hubble parameter. There are two sets of measurements of $H_0$ which are operative at high- and low-redshift patches. The early-time measurements are based on
CMB radiation or observations of baryonic acoustic oscillations (BAO) which give $H_0=67.36\pm 0.54~kms^{-1} Mpc^{-1}$ \cite{a}. The other set is based on late-time (or local) measurements such as Cepheids-Calibrated Supernovae which gives $H_0=74.03\pm 1.42~kms^{-1} Mpc^{-1}$ \cite{b}. These
two values are discrepant at $>4\sigma$. In absence of systematic errors \cite{c}, new physics or changing the expansion history seems to be necessary to remove the tension between the two sets of observations.\\
Some extensions of $\Lambda$CDM (either at early times or late times) have been proposed to resolve the problem \cite{d}. An early-time proposal consists of increasing the expansion rate around matter/radiation equality due to a new energy component known as early dark energy (EDE) \cite{e}. EDE behaves as a cosmological constant before some critical redshift $z_c$ while after that it dilutes faster than radiation to leave later expansion history unchanged. Another proposal relies on possible interactions between dark energy (DE) and dark matter (DM). In fact, different authors have already argued that due to current lack of knowledge of the nature of DM and DE it is quite possible that they interact and the interaction plays roles in expansion history of the Universe. For instance, interacting dark energy (IDE) models can offer explanations to the cosmic coincidence problem \cite{g} and the cosmological constant problem \cite{h}. It is shown \cite{f} that these interactions can also alleviate the current tension on the value of the Hubble constant $H_0$. \\
Just like quintessence, dynamical dark energy which gives accelerating expansion at late times, EDE is usually taken as a dynamical scalar field which can generate accelerating expansion near the recombination era. In order for EDE to have successful dynamics, it should arise without fine tuning and imposing special initial conditions. To avoid fine tuning, it is necessary to find attractor solutions in dynamics of EDE. If such attractor solutions exist the system evolves for a wide range of initial
conditions towards a unique asymptotic behavior. In the present work, we show that such attracting solutions actually exist when there is an interaction between DE and DM.
~~~~~~~~~~~~~~~~~~~~~~~~~~~~~~~~~~~~~~~~~~~~~~~~~~~~~~~~~~~~~~~~~~~~~~~~~~~~~~~~~~~~~~~~~~~~~~~~~~~~~~~~~~~~~~~~~~~~~~~~~~~~~~~~~~~~
\section{The Model}
We will consider a two-component interacting system consisting of DM and a real scalar field $\varphi$ as DE. Such an interaction is characterized by the following conservation equations
\begin{equation}
\nabla^{\mu}T_{\mu\nu}^{dm}=-\nabla^{\mu}T_{\mu\nu}^{\varphi}=Q_{\nu}
\label{0}\end{equation}
where $T_{\mu\nu}^{dm}$ and $T_{\mu\nu}^{de}\equiv T_{\mu\nu}^{\varphi}$ are energy-momentum tensors of DM and DE, respectively. The  interaction current $Q_{\nu}$ characterizes the energy transfer between the two energy components. Assuming a flat Friedmann-Robertson-Walker metric $ds^2=-dt^2+a^2(t)(\delta_{ij}dx^i dx^j)$ with $"a"$ being the scale factor, the energy balance equations (\ref{0}) take the form
\begin{equation}
\dot{\rho}_{\varphi}+3H(\omega_{\varphi}+1)\rho_{\varphi}=-Q
\label{a1}\end{equation}
\begin{equation}
\dot{\rho}_{dm}+3H\rho_{dm}=Q \label{a2}\end{equation}
where
$\rho_{dm}$ is energy density of DM, $\rho_{\varphi}\equiv \frac{1}{2}\dot{\varphi}^2+V(\varphi)$, $p_{\varphi}\equiv \frac{1}{2}\dot{\varphi}^2-V(\varphi)$, $\omega_{\varphi}\equiv p_{\varphi}/\rho_{\varphi}$ and $V(\varphi)$ is the potential function. The Hubble parameter $H\equiv\frac{\dot{a}}{a}$ satisfies the Friedmann equation\footnote{Hereafter we use the units $\kappa^2\equiv8\pi G=1$ where $G$ is the gravitational constant.}
\begin{equation}
3H^2=\rho_{dm}+\rho_{\varphi}
\label{a3}\end{equation}
The exact form of the coupling function $Q$ is unknown and a wide variety
of theoretical and phenomenological interacting scenarios
have been proposed and investigated in the literature \cite{1} \cite{2} \cite{3}. Here we consider the consequences of the interaction function\footnote{The covariant form of this function is $Q_{\nu}=\gamma T^{dm} \nabla_{\nu}\varphi$ where $T^{dm}$ is the trace of the energy-momentum tensor of dark matter. The trace $T^{dm}$ is solely given by $\rho_{dm}$ for the pressureless dark matter. The trace dependence of $Q_{\nu}$ means that any traceless energy component such as radiation does not contribute to the interacting system (\ref{a1}) and (\ref{a2}).} \cite{4}
\begin{equation}
Q=\gamma\rho_{dm}\dot{\varphi}
\label{3a}\end{equation}
This arises in different cases such as string theory, Einstein frame representation of scalar-tensor and $f(R)$ gravity theories.
Although $\gamma$ is generally an evolving function and depends on several variables, we will take it as a constant parameter. This is a good approximation if the coupling function $\gamma(\varphi,...)$ is slowly varying or the process of interaction takes place in a short period of time with respect to the whole expansion history of the Universe. The latter is particularly true for EDE scenario. \\
The equation (\ref{a2}) can be immediately solved to give
\begin{equation}
\rho_{dm}=\rho_{0dm} a^{-3} e^{\gamma\varphi}
\label{4}\end{equation}
with $\rho_{0dm}$ being an integration constant. This equation is equivalent to
$\rho_{dm}=\rho_{0dm}~a^{(-3+\varepsilon)}$
with $\varepsilon=\frac{\gamma\varphi}{\ln a}$. While the coupling function $Q$ represents the interaction of DM and DE, $\varepsilon$ indicates how the interaction affects the expansion rate of the Universe. Note that if $\varepsilon$ is taken to be a constant parameter then $\dot{\varphi}=\frac{\varepsilon}{\gamma}H$. In this case, (\ref{3a}) would be equivalent to $Q=\varepsilon\rho_{dm}H$. This is the coupling function of interacting models discussed by some authors \cite{int}. \\
In the case of null interaction ($Q=0$), DM energy density follows $\rho_{dm}=\rho_{0dm}~a^{-3}$, the standard matter evolution in a matter-dominated Universe. On the other hand, $Q\neq 0$ corresponds to an energy exchange between DM and DE. In this case when
$\varepsilon>0$ ($\gamma>0$), energy is injecting
into DM so that the latter will dilute more slowly compared
to the standard evolution. When $\varepsilon<0$ ($\gamma<0$) the reverse is true, namely that DM is annihilated and the rate of its dilution is faster than
the case $\rho_{dm}\propto a^{-3}$.\\
~~~~~~~~~~~~~~~~~~~~~~~~~~~~~~~~~~~~~~~~~~~~~~~~~~~~~~~~~~~~~~~~~~~~~~~~~~~~~~~~~~~~~~~~~~~~~~~~~~~~~~~~~~~~~~~~~~~~~~~~
\section{Phase space analysis}
To study dynamics of the equations (\ref{a1}), (\ref{a2}) and (\ref{a3}) one should write them as a system of autonomous ordinary differential
equations by defining a set of appropriate variables. In general, there are many possible ways to define these variables.
We will use \cite{cop},
\begin{equation}
X=\frac{1}{H}\frac{\dot{\varphi}}{\sqrt{6}}~~~~,~~~Y=\frac{1}{H}\sqrt{\frac{V}{3}}
\label{5}\end{equation}
We also consider an exponential potential $V(\varphi)=e^{-\lambda\varphi}$ \cite{poten} where $\lambda$ is a constant parameter.  In terms of (\ref{5}), the constraint (\ref{a3}) takes the form
\begin{equation}
1=\Omega_{dm}+\Omega_{\varphi}=\Omega_{dm}+X^2+Y^2
\label{5a}\end{equation}
where $\Omega_{dm}=\rho_{dm}/3H^2$ and $\Omega_{\varphi}=\rho_{\varphi}/3H^2=X^2+Y^2$ are the density parameters of DM and DE, respectively. Hence the two-dimensional phase space is restricted to the region $0\leq X^2+Y^2 \leq 1$ and $Y\geq 0$. This means that the physical phase space of
the model is contained within the unit circle. The deceleration parameter and the effective equation of state of the two interacting fluids are given by
\begin{equation}
q=-1-\frac{\dot{H}}{H^2}=\frac{1}{2}+\frac{3}{2}(X^2-Y^2)
\label{10q}\end{equation}
\begin{equation}
\omega_{eff}=-1-\frac{2\dot{H}}{3H^2}=X^2-Y^2
\label{10}\end{equation}
Now the set of equations (\ref{a1}), (\ref{a2}) and (\ref{a3}) with variables (\ref{5}) gives rise to the autonomous equations
\begin{equation}
X'=-3X+\sqrt{\frac{3}{2}}\lambda Y^2-\sqrt{\frac{3}{2}}\gamma(1-X^2-Y^2)-X\frac{H'}{H}
\label{6}\end{equation}
\begin{equation}
Y'=-\sqrt{\frac{3}{2}}\lambda X Y -Y\frac{H'}{H}
\label{7}\end{equation}
where the prime indicates derivative with respect to $\ln a$. The expression for $\frac{H'}{H}$ can be straightforwardly obtained by taking derivative of (\ref{a3}) and using (\ref{a1}) and (\ref{a2}). It gives
\begin{equation}
\frac{H'}{H}=-\frac{3}{2}(1+X^2-Y^2)
\label{7a}\end{equation}
Combining the latter with (\ref{6}) and (\ref{7}) leads finally to
\begin{equation}
X'=f_X(X, Y)\equiv-3X+\sqrt{\frac{3}{2}}\lambda Y^2-\sqrt{\frac{3}{2}}\gamma(1-X^2-Y^2)+\frac{3X}{2}(1+X^2-Y^2)
\label{8}\end{equation}
\begin{equation}
Y'=f_Y(X, Y)\equiv-\sqrt{\frac{3}{2}}\lambda X Y +\frac{3Y}{2}(1+X^2-Y^2)
\label{9}\end{equation}
The fixed points are those that satisfy $X'=0$ and $Y'=0$.  These points together with $\omega_{eff}$, $\Omega_{\varphi}$ and the deceleration parameter are given in Tab.1 for the system (\ref{6}) and (\ref{7}).
\begin{table}[h]
\caption{The fixed points and the corresponding effective equation of state for the system (\ref{8}) and (\ref{9}) }
\centering
\begin{tabular}{|c| c| c| c| c|c|c|}
\hline\hline

Name & X   & Y   & $\omega_{eff}$ & $\Omega_{\varphi}$ & $q$ & $q<0$   \\

 \hline

$A_{+}$ & $1$ & $0$ & $1$ & $1$ & $2$ & $no$ \\
\hline

$A_{-}$ &  $-1$ & $0$ & $1$ & $1$ & $2$ & $no$   \\

\hline

B & $-\sqrt{\frac{2}{3}}\gamma$ &$0$ & $\frac{2}{3}\gamma^2$ &  $\frac{2}{3}\gamma^2$ & $\frac{1}{2}+\gamma^2$ & $no$                \\

\hline

$C_{+}$ & $\frac{\lambda}{\sqrt{6}}$ &  $\sqrt{1-\frac{\lambda^2}{6}}$ & $-1+\frac{\lambda^2}{3}$  & $1$ & $-1+\frac{\lambda^2}{3}$ & $\lambda<\sqrt{2}$ \\
\hline
$C_{-}$ & $\frac{\lambda}{\sqrt{6}}$ &  $-\sqrt{1-\frac{\lambda^2}{6}}$ & $-1+\frac{\lambda^2}{3}$ & $1$ & $-1+\frac{\lambda^2}{3}$ & $\lambda<\sqrt{2}$   \\
\hline
$D_{+}$ &$\frac{\sqrt{\frac{3}{2}}}{(\gamma+\lambda)}$  &$\sqrt{\frac{2\gamma(\gamma+\lambda)+3}{2(\gamma+\lambda)^2}}$ & $-\frac{\gamma}{(\gamma+\lambda)}$  & $\frac{\gamma(\gamma+\lambda)+3}{(\gamma+\lambda)^2}$ & $\frac{1}{2}-\frac{3\gamma}{2(\gamma+\lambda)}$ & $\frac{\gamma}{(\gamma+\lambda)}> \frac{1}{3}$  \\

\hline
$D_{-}$ &$\frac{\sqrt{\frac{3}{2}}}{(\gamma+\lambda)}$  &$-\sqrt{\frac{2\gamma(\gamma+\lambda)+3}{2(\gamma+\lambda)^2}}$ & $-\frac{\gamma}{(\gamma+\lambda)}$ & $\frac{\gamma(\gamma+\lambda)+3}{(\gamma+\lambda)^2}$ & $\frac{1}{2}-\frac{3\gamma}{2(\gamma+\lambda)}$ & $\frac{\gamma}{(\gamma+\lambda)}> \frac{1}{3}$     \\
\hline

\end{tabular}
\label{1}
\end{table}
\begin{table}[h]
\caption{The fixed points and the corresponding effective equation of state for the system (\ref{8}) and (\ref{9}) }
\centering
\begin{tabular}{|c| c| c| c| }
\hline\hline

point & $\mu_1$  & $\mu_2$   & $stability$    \\

 \hline

$A_{+}$ & $3+2\sqrt{\frac{3}{2}}\gamma$ & $3-\sqrt{\frac{3}{2}}\lambda$ & $\gamma<-\frac{3}{2}$, $\lambda>\sqrt{6}$ \\
\hline

$A_{-}$ &  $3-2\sqrt{\frac{3}{2}}\gamma$ & $3+\sqrt{\frac{3}{2}}\lambda$ & $\gamma>\frac{3}{2}$, $\lambda<-\sqrt{6}$   \\

\hline

B & $-\frac{3}{2}+\gamma^2$ &$\frac{3}{2}+\gamma(\lambda+\gamma)$ & $\gamma<\pm \sqrt{\frac{3}{2}}$ , $\gamma(\lambda+\gamma)<-\frac{3}{2}$                \\

\hline

$C_{\pm}$ & $\frac{\lambda}{2}(\lambda^2-6)$ &  $-3+\lambda(\lambda+\gamma)$ & $\{\lambda<-\sqrt{6}\}\bigcup \{0<\lambda<\sqrt{6}\}$, $\lambda(\lambda+\gamma)<3$  \\
\hline

$D_{+}$ &$$  &$$ & $Fig.1$    \\

\hline
$D_{-}$ &$$  &$$ & $Fig.1$    \\
\hline

\end{tabular}
\label{1}
\end{table}
In order to study stability
of the fixed points, one should consider perturbations around them. In the linearized case, $\partial f_i/\partial x_j$ is of particular importance. In our two-dimensional case, the eigenvalues (denoted in the following by $\mu_1$ and $\mu_2$) of the Jacobian matrix
\[
  J=
  \left[ {\begin{array}{cc}
   \partial f_X/X & \partial f_X/Y \\
   \partial f_Y/X & \partial f_Y/Y \\
  \end{array} } \right]
\]
determine stability of the fixed points. The fixed points are stable if both eigenvalues are
negative, unstable if both are positive and saddle
if one is positive and the other is negative.
For the dynamical system (\ref{6}) and (\ref{7}), eigenvalues of $J$ and the stability status of the fixed points are given in Tabs.2. The points $D_{\pm}$ have more complicated stability conditions. The stability status of $D_{+}$ is depicted in Fig.1. The figure shows that $D_{+}$ is stable ($\mu_{1,2}<0$) or saddle ($\mu_{1}<0,~\mu_{1}>0$ and vice versa) in most regions of the parameters space.\\
The fixed points $A_{\pm}$ represent solutions for which the kinetic energy of DE is dominated ($X=\pm 1$, $Y=0$). For these solutions, the right hand side of (\ref{a3}) corresponds to an effective fluid with a stiff equation of state $\omega_{eff}=1$. The point $B$ is generalization of the unstable solution $x=0,~y=0$ of \cite{cop} in the non-interacting case $\gamma=0$. In our case, this point is DM-dominated (for $\gamma<1$) which is affected by the interaction and can now be stable for some values of $\gamma$ and $\lambda$ (see Tab. 2).
The stream plot of the system (\ref{8}) and (\ref{9}) is indicated in Fig. 2. In Fig.2(a), $A_{\pm}$ are unstable nodes for which all the trajectories evolve away from them and then are attracted towards the late-time attractor $C_{+}$. In Fig.2(b),(c), all trajectories emerge from the early-time attractor $A_{-}$ before being attracted towards the saddle point $A_{+}$. These trajectories ultimately end to the late-time attractors $B$ and $D_{+}$.

~~~~~~~~~~~~~~~~~~~~~~~~~~~~~~~~~~~~~~~~~~~~~~~~~~~~~~~~~~~~~~~~~~~~~~~~~~~~~~~~~~~~~~~~~~~~~~~~~~~~~~~~~~~~

\section{Conclusions}
Dynamical dark energy models or quintessence are based on scalar fields with appropriate potentials which can produce a late-time accelerating expansion. The same scenario is recently used to alleviate the tension between local determination of $H_0$ and the values inferred by CMB experiments at recombination era. In fact, EDE is a scalar field which is responsible for the required accelerating expansion at early times to alleviate the Hubble tension. It behaves as a cosmological constant just before recombination and then dilutes away like radiation or faster at late times. 
Due to the unknown nature of the dark sector it is quite possible that there is an interaction between DM and DE both at early and late times. At late times, it is shown that interacting dark energy models play roles for addressing some cosmological puzzles such as the coincidence and the cosmological constant problems.\\ In this Letter, the main focus is  on the problem of initial conditions.  We consider a cosmological model in which DM and DE are allowed to have mutual interaction via the coupling function $Q=\gamma\rho_{dm}\dot{\varphi}$. The advantage of this choice is that it arises from Einstein frame representation of scalar-tensor theories and $f(R)$ gravity models. We first connect this model to the class of models characterized by the coupling function $Q\propto \rho_{dm}H$. This connection is established by taking $\varepsilon$, denoting how the rate of dilution of $\rho_{dm}$ differs from $\rho_{dm}\propto a^{-3}$, as a constant parameter. Then we have used a dynamical analysis approach and an exponential potential for the DE scalar field to show that there are early- and late-time attractor solutions in such a dynamical system. This ensures that evolution of trajectories of the system converges to a unique asymptotic behavior for a much wider range of initial conditions. Moreover, we have shown that a DE-dominated ($\Omega_{\varphi}=1$) solution exists which is a late-time attractor (Fig. 1(a)). We have also found different late-time attractors
where neither DE nor DM dominates the evolution of the Universe.

~~~~~~~~~~~~~~~~~~~~~~~~~~~~~~~~~~~~~~~~~~~~~~~~~~~~~~~~~~~~~~~~~~~~~~~~~~~~~~~~~~~~~~~

\begin{figure}[h]
\begin{center}
\includegraphics[width=0.4\linewidth]{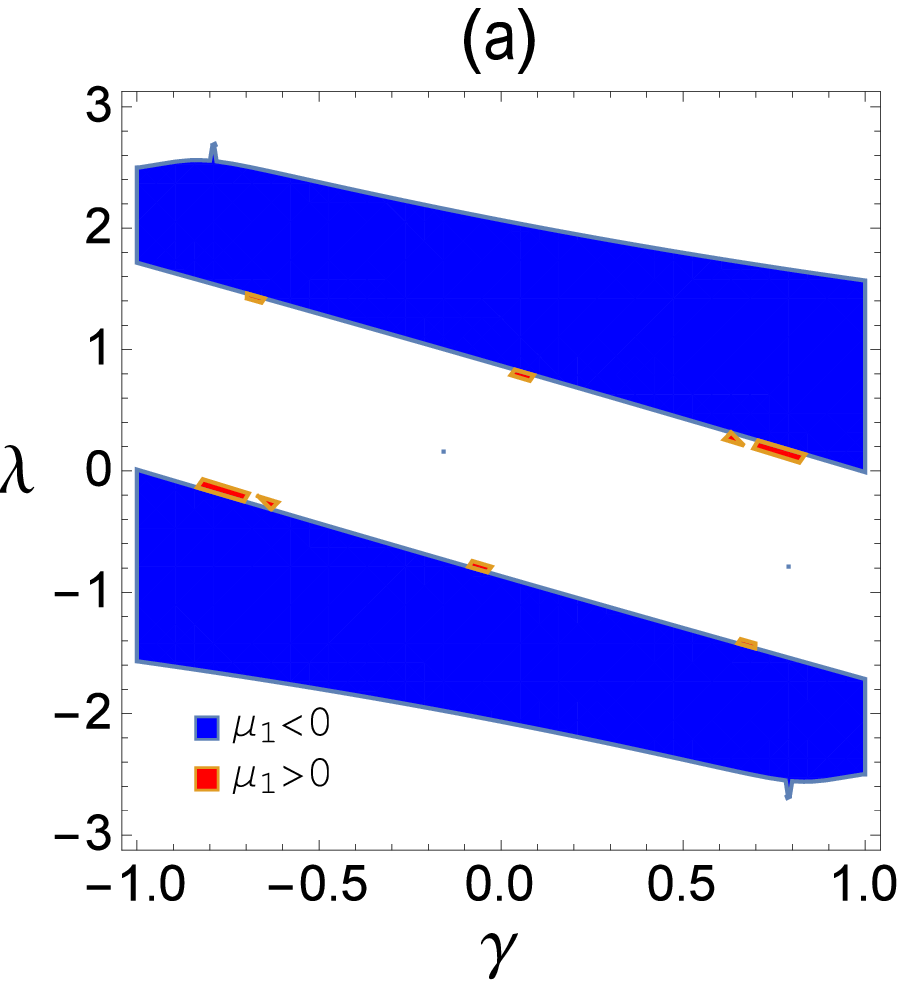}
\includegraphics[width=0.4\linewidth]{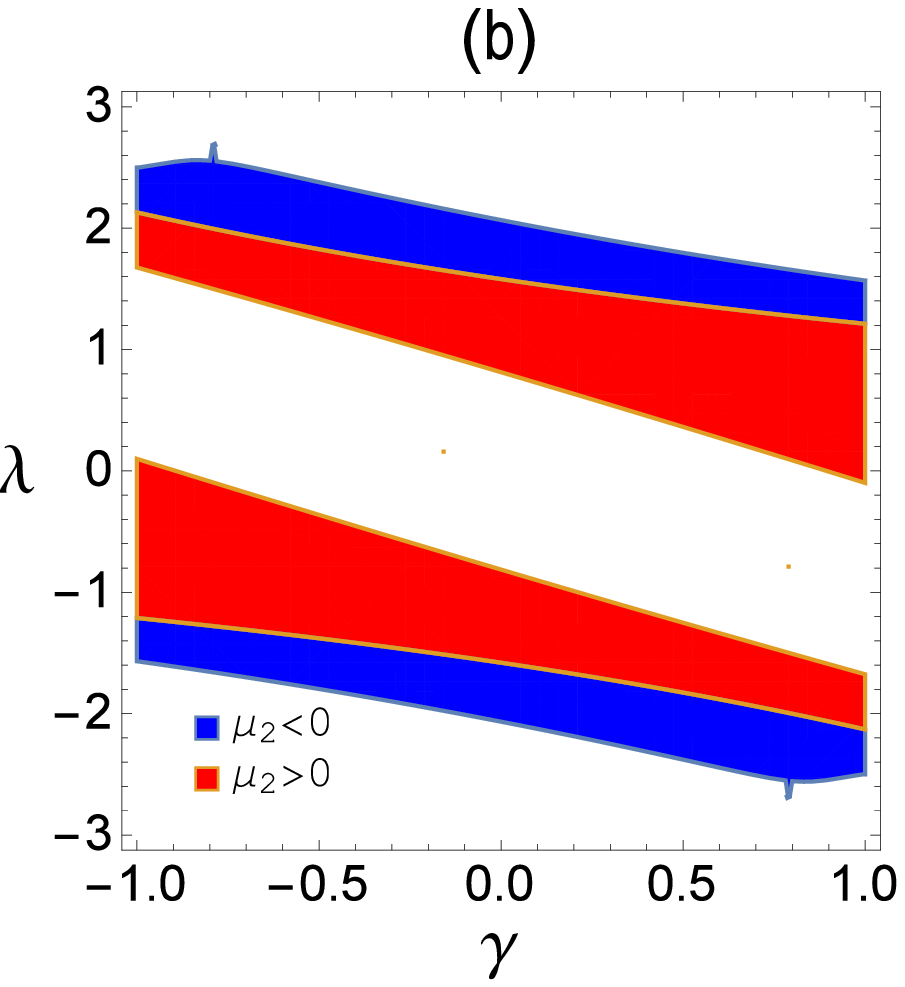}
\caption{The parameters spaces corresponding to the eigenvalues (a) $\mu_1$ and (b) $\mu_2$ of the point $D_{+}$.}
\end{center}
\end{figure}
\begin{figure}[h]
\begin{center}
\includegraphics[width=0.4\linewidth]{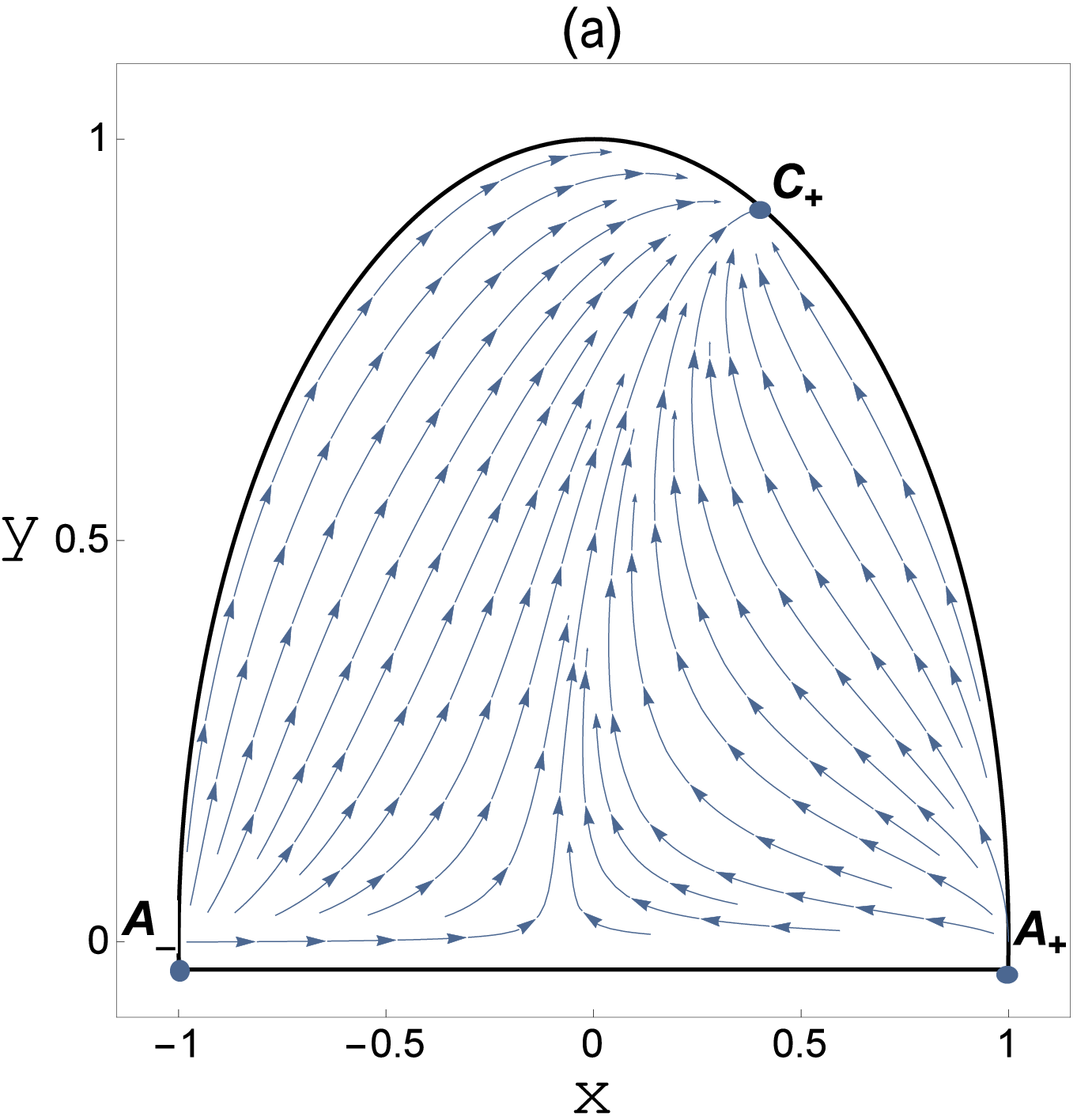}
\includegraphics[width=0.4\linewidth]{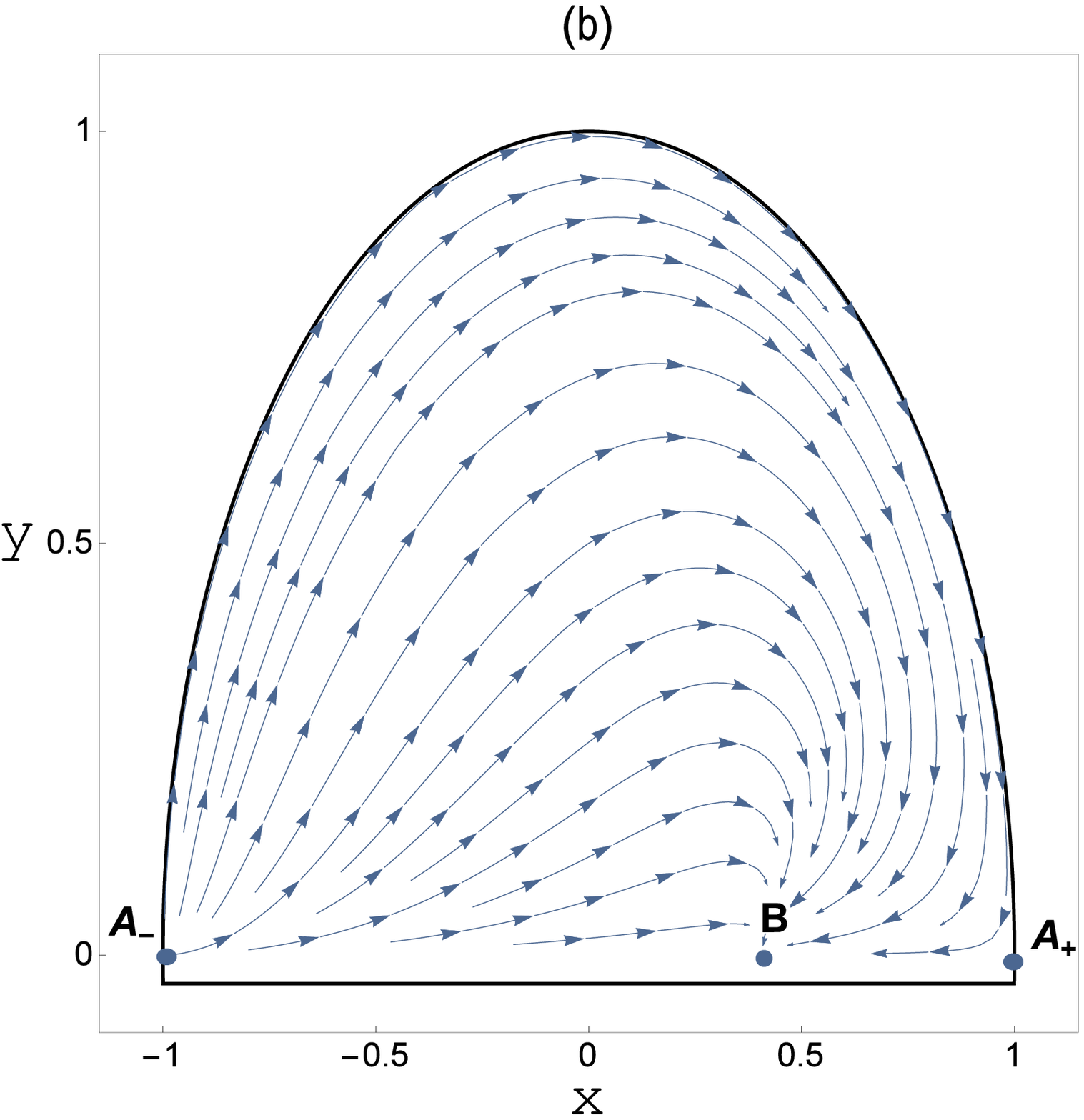}
\includegraphics[width=0.4\linewidth]{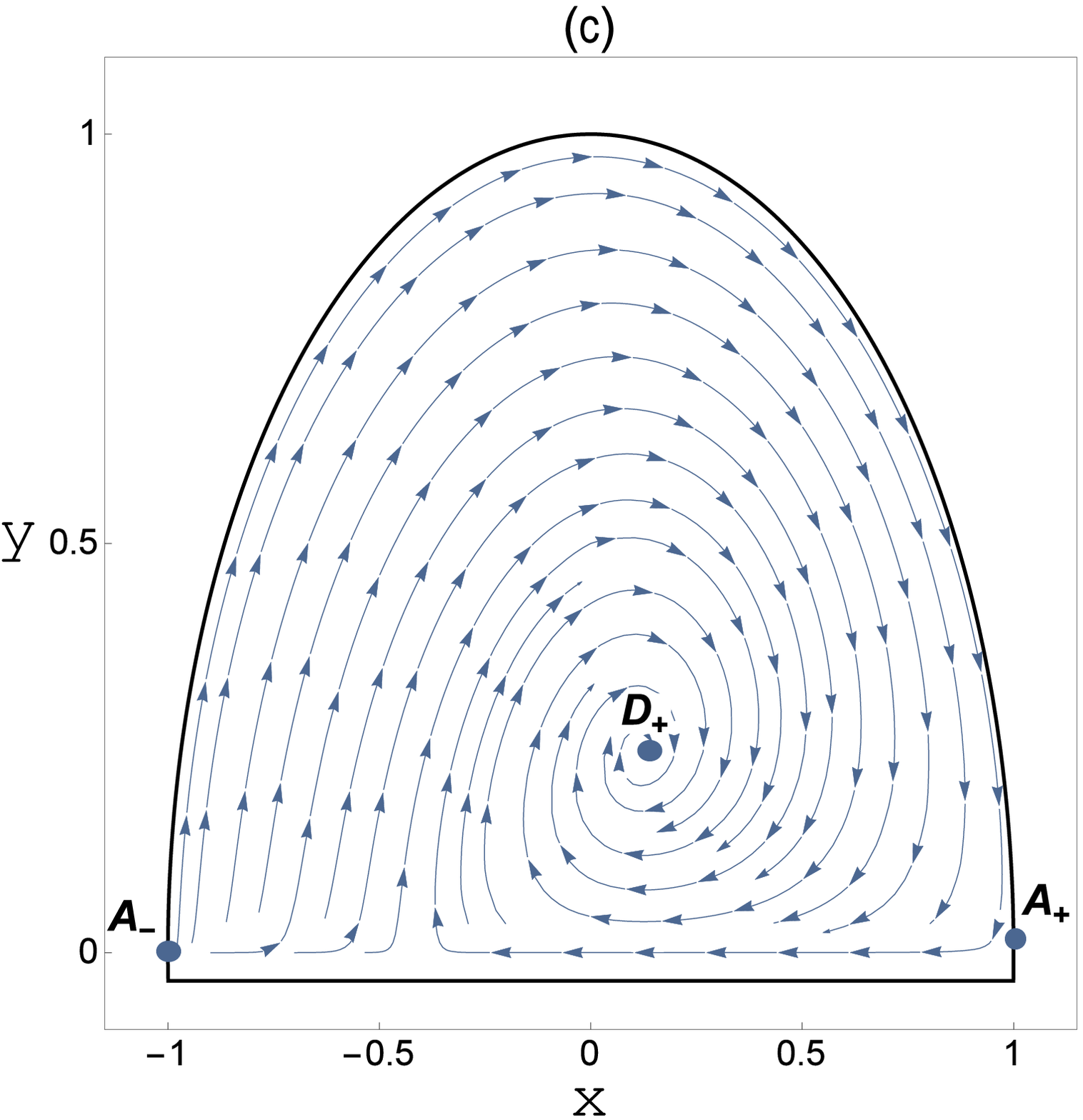}
\caption{The phase plots of the dynamical system (\ref{8}) and (\ref{9}) for a) $\lambda=1,~ \gamma=0.1$ , b) $\lambda=4,~ \gamma=-0.5$ and c) $\lambda=10,~ \gamma=0.5$ .}
\end{center}
\end{figure}

\end{document}